\documentclass[conference]{IEEEtran}
\IEEEoverridecommandlockouts

\usepackage{amsmath}
\usepackage{amssymb}
\usepackage{mathrsfs}
\usepackage{amsfonts}
\usepackage{array}
\usepackage{tabu}

\usepackage{listings}

\usepackage{balance}

\usepackage{cite}

\usepackage{ifpdf}

\usepackage{multicol}
\usepackage{makeidx}
\usepackage[usenames]{color}
\usepackage{soul}
\usepackage{pseudocode}
\usepackage{epsf}

\newcommand{\be}{\begin{eqnarray}}
\newcommand{\ee}{\end{eqnarray}}

\ifCLASSINFOpdf
   \usepackage[pdftex]{graphicx}
   \graphicspath{{./fig/pdf/}}
   \DeclareGraphicsExtensions{{.pdf,.png}}
\else
   \usepackage[dvips]{graphicx}
   \graphicspath{{./fig/eps/}}
   \DeclareGraphicsExtensions{{.eps,.png}}
\fi

\def\refname{\normalsize \centerline{\bf REFERENCES} }
\def\thefootnote{\fnsymbol{footnote}}

\DeclareMathAlphabet{\mathpzc}{OT1}{pzc}{m}{it}

\usepackage{caption}
\usepackage[labelformat=simple]{subcaption}

\begin{document}

\title{Biometrics-as-a-Service: A Framework to Promote Innovative Biometric Recognition in the Cloud
 \vspace{-0.15cm} }

\author{\IEEEauthorblockN{
Veeru Talreja,\IEEEauthorrefmark{1}
Terry Ferrett,\IEEEauthorrefmark{1}
Matthew C. Valenti,\IEEEauthorrefmark{1}
Arun Ross\IEEEauthorrefmark{2}
}
\IEEEauthorblockA{
\IEEEauthorrefmark{1}West Virginia University, Morgantown, WV, USA.\\
\IEEEauthorrefmark{2}Michigan State University, East Lansing, MI, USA.}
\vspace{-0.35cm}
}

\maketitle

\begin{abstract}

Biometric recognition, or simply \emph{biometrics}, is the use of biological attributes such as face, fingerprints or iris in order to recognize an individual in an automated manner. A key application of biometrics is authentication; i.e., using said biological attributes to provide access by verifying the claimed identity of an individual. This paper presents a framework for \emph{Biometrics-as-a-Service} (BaaS) that performs biometric matching operations in the cloud, while relying on simple and ubiquitous consumer devices such as smartphones. Further, the framework promotes innovation by providing interfaces for a plurality of software developers to upload their matching algorithms to the cloud. When a biometric authentication request is submitted, the system uses a criteria to automatically select an appropriate matching algorithm. Every time a particular algorithm is selected, the corresponding developer is rendered a micropayment. This creates an innovative and competitive ecosystem that benefits both software developers and the consumers. As a case study, we have implemented the following: (a) an ocular recognition system using a mobile web interface providing user access to a biometric authentication service, and (b) a Linux-based virtual machine environment used by software developers for algorithm development and submission.  

\end{abstract}

\IEEEpeerreviewmaketitle

\section{Introduction}

\emph{Biometrics} is the science of recognizing individuals based on their physical or behavioral attributes such as face, fingerprints, iris, gait, or voice\cite{jain_intro_2004}. Biometric systems are applied in a number of domains including law enforcement, access control, healthcare, and financial services. The primary function of a biometric system is to compare two biometric samples and determine whether they pertain to the same individual or different individuals by generating a match score. Thus, a classical biometric system obtains a sample of the user's attribute via \emph{sensing}, processes the sample to extract a discriminative feature set via \emph{feature extraction}, and \emph{compares} the extracted feature set with those in a database in order to verify a user's identity (``verification") or to determine a user's identity (``identification"). The database consists of a list of user identities (e.g., names or numbers), with each identity being linked to a feature set extracted from the user's biometric sample during enrollment.

More recently, there has been tremendous interest in incorporating biometric solutions into consumer electronic such as {\em smartphones}. Traditionally, biometric sensors have been proprietary with high cost, low market adoption, and limited compatibility with competing systems \cite{rose2016biometrics}.  However, smartphones are equipped with cameras and other sensors suitable for biometric sensing tasks in the context of face, fingerprint, ocular and gait recognition. The presence of high-resolution cameras, in particular, offers the possibility of performing face or periocular recognition within the confines of the device. Modalities in addition to iris for ocular biometrics have gained considerable research attention in the recent past \cite{woodard2010fusion,alonso2015near,raghavendra2016learning}.

Face or ocular recognition using mobile cameras is challenging due to several confounding factors including variations in head pose, ambient illumination, facial expression and occlusion \cite{JillelaRossMobile2015,freitas_periocmob_2015}. Another significant challenge is the limited availability of resources - within the smartphone - for storage and computation. Therefore, it may be necessary in some cases to outsource the computing and/or storage demands to a more powerful server outside the smartphone. In this regard, {\em cloud computing} may be harnessed as a viable option. Cloud computing \cite{nist_cloud_2011} facilitates the outsourcing of computing and storage tasks to infrastructures managed by dedicated providers \textemdash  a potential approach to surpassing mobile resource limits \cite{stojmenovic_cloud_2012}. For instance, the feature extraction, data storage, and matching components of a biometric system can be moved to a cloud infrastructure, while leaving only the sensing task in the smartphone. This is an example of a \emph{biometrics-in-the-cloud} architecture \cite{das_biometrics_2013}.

There is an increased interest in performing biometric recognition in mobile devices and as a cloud-based service \cite{jeong_irismob_2006,chow_auth_2010,kang_mobiris_2010,barra_irismob_2015,bharadi_sigrec_2015}.  In \cite{bommagani_framebio_2014}, a framework for cloud-based face recognition emphasizing the parallelization of recognition tasks across multiple servers is introduced.

 If the biometrics-in-the-cloud architecture is offered by a service provider, then it is referred to as \emph{Biometrics-as-a-Service} (BaaS) \cite{rose2016biometrics}. If the infrastructure allows for component developers to develop and incorporate custom components in the cloud (e.g., feature extraction or matcher modules), then it is referred to as \emph{Platform-as-a-Service} (PaaS). Some PaaS providers, such as Bungee Labs and SalesForce.com  \cite{lawton2008developing}, provide a framework that allows independent software vendors (ISV) to develop extensions or add-ons to the provider's core application \cite{beimborn2011platform}.  A key contribution of this paper is to establish a similar framework that allows the developers of biometric-recognition algorithms to actively contribute to the BaaS system.  This is achieved by creating an interface for uploading algorithms and a scheme for selecting algorithms and rendering micropayments.  Having such an infrastructure in place has the benefit of promoting innovation and reducing costs for the BaaS by allowing the development of its key components to be outsourced.  

  In this paper, we propose a cloud-based framework for performing {\em biometric recognition} using smartphones as sensors and demonstrate an implementation of this framework for ocular recognition. The salient features of this framework include the following:
\begin{enumerate}
\item Smartphones, including the cameras resident in them, require no modification from their stock hardware configuration. 
\item Computationally intensive tasks such as segmentation, feature extraction and matching are outsourced to the cloud. 
\item Software developers can upload their own biometric matching algorithms to the cloud. Thus, the cloud hosts multiple matching algorithms pertaining to multiple developers. 
\item When two biometric images are submitted to the cloud, the matching algorithm is automatically selected based on the characteristics of the input images and the performance of the algorithm as evaluated on sequestered data. 
\item Every time an algorithm is selected for matching, its developer is credited under a \emph{micropayment} model.
\end{enumerate}

Enabling developers to upload matching algorithms creates an environment where the value of algorithms is measured by their in-application performance, creating incentives for competition and innovation.

\section{Architecture}\label{sec:arch}

\begin{figure}[t]
\centering
\includegraphics[width=9 cm]{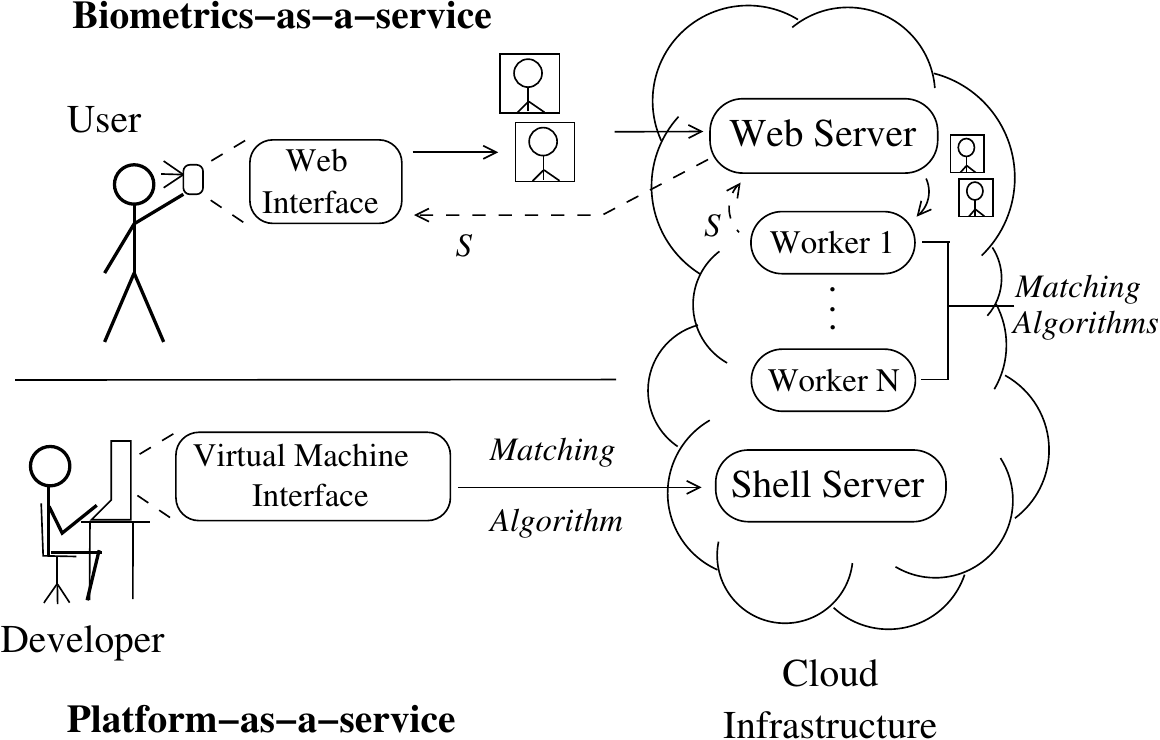}
\caption{Proposed BaaS/PaaS architecture.  A user submits a matching request by uploading two images captured using a smartphone camera to the web interface. The comparison is performed by a worker process and a matching score, $S$ is returned. Matching algorithm developers use the virtual machine interface to develop and submit their algorithms to the cloud infrastructure over a shell session for deployment to the worker processes.}\label{fig:arch} 
\vspace{-0.40cm}
\end{figure}

This section develops a general framework for BaaS and PaaS using smartphones, 
focusing on the iris and periocular modalities as examples (together referred to as ``ocular" in this paper). The components of the system are the \emph{user interface}, \emph{developer interface},
and cloud-based \emph{computing infrastructure}.
The user interface is a mobile-enabled web application through which users submit
matching requests (i.e., images to be compared) and receive results (i.e., match scores).
The developer interface is a virtual machine having identical software as the
computing infrastructure for developing and submitting matching algorithms to
the system.
The computing infrastructure consists of a cloud server for executing
matching requests using developer submitted algorithms.
The architecture is depicted in Fig. \ref{fig:arch}.

\subsection{Cloud Computing Characteristics}

The definition of cloud computing \cite{nist_cloud_2011} encompasses several elements
which our system provides.  Users can submit matching requests through a web interface,
and the requests are automatically processed by servers as available, providing \emph{on-demand self-service}.
The web interface is designed for both mobile and desktop use, incorporating \emph{broad network access}.
\emph{Resource pooling} is implemented such that multiple matching requests are distributed among servers, automatically balancing the request load as needed.

Additional servers can be added to the system rapidly, as operating system 
installation and software provisioning are fully automated, providing \emph{rapid elasticity}.
The computing framework software is designed to execute matching
requests across any number of servers.
The execution of each matching request is tracked,
and users are charged for service in proportion to the number of completed requests. 
Matching algorithm developers are credited for every matching request which uses
their algorithm, making this an instance of \emph{measured service}.

Cloud services are classified according to the level of abstraction at
which the users and developers interact with the infrastructure.  
In the present work, users perform matching operations by submitting
requests through a web interface. In the context of biometrics, this architecture
is an instance of \emph{biometrics-as-a-service}, a model for providing
biometric recognition functionality through a service provider \cite{rose2016biometrics}.
A virtual machine containing an operating system and pre-installed software that 
is identical to the cloud infrastructure is provided to software
developers, making this an instance of \emph{platform-as-a-service}. 
This virtual machine ensures that developed algorithms are binary-compatible with
the infrastructure, and obviates the need for developers to provision their own development environments.

\subsection{User Interface}

The \emph{user interface} is a web application which is accessible on both mobile and desktop browsers.
This interface is used by the user to submit two ocular images for matching and to receive the result. These uploaded images are sent to the cloud infrastructure where preprocessing is done to select the most suitable matching algorithm. The selected algorithm is executed on the images and a match score is returned to the user.

\subsection{Developer Interface}\label{subsec:dev_int_model}

The \emph{developer interface} is a \emph{virtual machine} (VM) containing software for developing
algorithms for submission to the cloud infrastructure.
The operating system and software environment on the VM is configured identically 
to the environment on the computing infrastructure (i.e., the cloud server).
Identical configuration obviates the need for the algorithm developer to invest time installing and configuring
a compatible development environment.
The VM enables the developer to implement matching algorithms which are binary-compatible
with the infrastructure, and upload scripts and executables directly for use.

The VM is distributed over the internet as an archive containing a disk image 
of the pre-installed and configured operating system.
The \emph{software hypervisor} which executes the virtual machine is chosen for compatibility
with as many widely-used host operating systems as possible.
A desktop environment having minimal resource requirements is chosen for the VM to 
enhance user interface performance in a virtualized environment.

The VM contains software for algorithm development, such as compilers, text editors, debuggers,
and source control clients, as well as libraries commonly used for image processing
applications and research.
The developer must implement their algorithm such that it can be executed on a command line 
- a broad and general requirement which is straightforward to satisfy.

\subsection{Computing Infrastructure}\label{subsec:cloudinf_mdl}

This subsection describes the computing infrastructure
which executes matching requests using developer-submitted
algorithms.
The data submitted to the web interface by the user is described, followed
by processing to schedule the matching request for execution.
The execution of the request is discussed, along with the data flow through which
the matching request is returned to the user via the web interface.

The user submits a matching request to the web interface by specifying two ocular images
and, optionally, a matching algorithm. 
If the user does not select an algorithm, the system selects one automatically.  
The web interface stores the images in the user's \emph{data directory} and creates
a \emph{job file} in the user's \emph{job input queue} containing parameters required for matching: filesystem paths to
the eye images and the matching algorithm, if an algorithm has been selected.

The \emph{job manager} preprocesses the job file and moves it to the user's \emph{job running queue}.
Preprocessing is performed as follows.
If the user selected a specific matching algorithm, no action is taken during preprocessing.
If no algorithm was selected by the user, the radii of the irises in the input image is calculated.
If the smaller of the two radii falls below a threshold, a general \emph{ocular} matching
algorithm is selected, otherwise an \emph{iris-based} matching algorithm is selected.
A table of matching algorithms and the number of times each has been executed is updated
by the job manager, increasing the number of executions for the selected algorithm by one.
Following preprocessing, the job manager creates a \emph{task file} containing the paths to the input eye images
and the selected matching algorithm and places the task file in the \emph{user task input queue}.

The \emph{task controller} determines the number of tasks contained in all users'
task input queues, and schedules tasks for execution such that all users receive
an equal share of the available processing cores, denoted as \emph{fair scheduling}.
The system is implemented such that other scheduling policies may be incorporated.
To schedule an individual task for execution, the task controller moves the task file 
from the user's task input queue to the \emph{global task input queue}.

A \emph{generic worker process} running on one of the worker nodes reads a task in
the global task input queue and moves it to the \emph{global task running queue}.
The generic worker process executes the matching operation, saves the 
result in the task file, and moves it to the \emph{global task output queue}.
The task controller moves the task from from the global task output queue to the 
\emph{user task output queue}.  
The task file is read by the job manager, the matching result is stored in 
the job file and moved to the \emph{job output queue}.  
The web interface reads the matching result from the job file and displays 
it to the user.

\section{Implementation}\label{sec:impl}

This section describes our implementation of the BaaS/PaaS architecture described in Section \ref{sec:arch}.
The user interface implementation is first described, followed by the developer interface.
Finally, details for the computing infrastructure implementation are given.

\subsection{User Interface}
\renewcommand{\thefootnote}{\arabic{footnote}}
In this section we define the specific implementation of the \emph{User Interface} along with an overview of the components used to build the interface. We used the Mobile-Google Web Toolkit (MGWT)\footnote[1]{http://www.m-gwt.com/} which is a software framework for developing mobile web applications. MGWT is an extension of Google Web Toolkit (GWT)\footnote[2]{http://www.gwtproject.org/}, which is  a Java based framework, for creating efficient and optimized browser based applications. GWT is an open-source completely free framework that helps developers to build high performance web applications without having expert skills in JavaScripting or browser quirks. Google also uses GWT in many of its products such as Adwords, and AdSense.  

While GWT can help build fast desktop applications using Java, it lacks widgets and animations for developing great mobile apps. MGWT closes this gap -- MGWT provides mobile widgets, smooth animations, touch support and much more. One can use MGWT to build highly optimized Java based AJAX applications that are compatible with all browsers, including Android and the iPhone mobile browsers. We used MGWT 1.1.2 along with GWT 2.7 and Eclipse for developing the user interface. A few other Java based API's were also used in sync with MGWT to develop the functionality of the user interface.    

Given below is a summary of steps required to go through the process of using the web interface for logging in, uploading two images and obtaining the matching score based on automatic or manual selection of the algorithm.

1. The web application can be accessed at \emph{https://wcrl.csee.wvu.edu/IrisCloud}.

2. After logging into the application, the user can either view their previous job submissions or submit a new job which entails uploading images and obtaining the match score.

\begin{figure}[t]
\centering
\begin{subfigure}[t]{0.22\textwidth}
                \includegraphics[width=\linewidth]{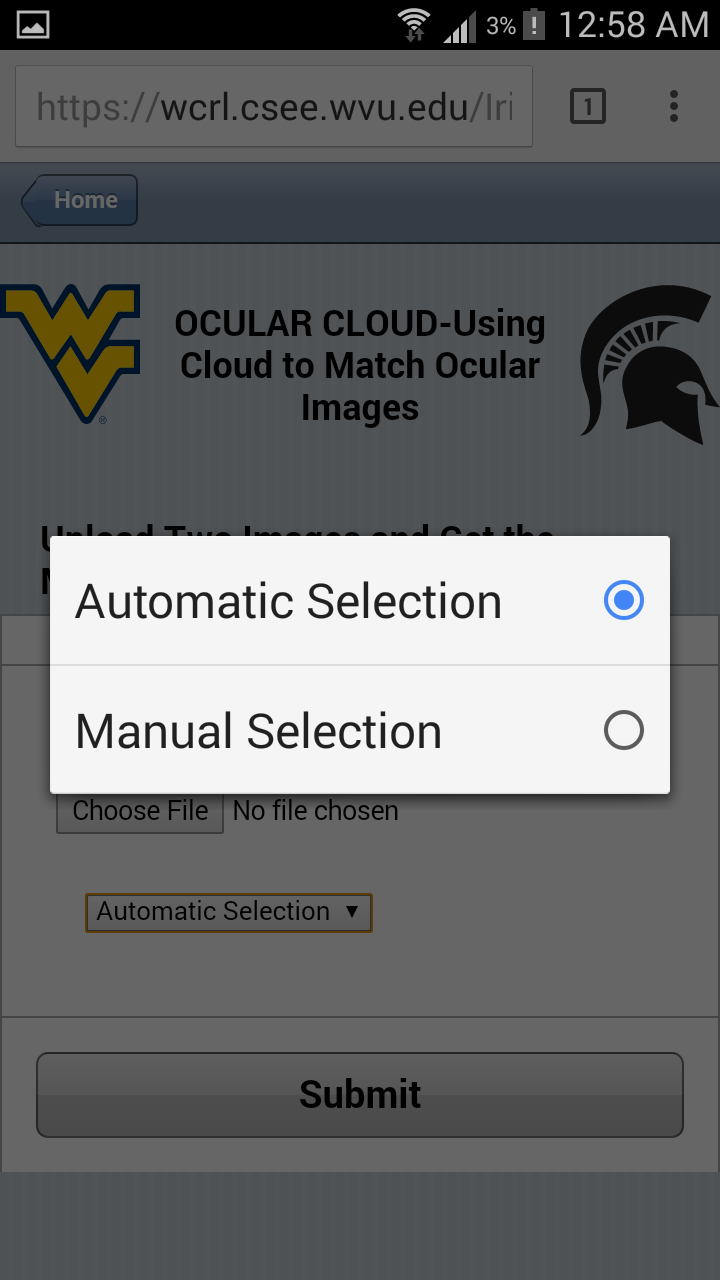}
                \caption{Algorithm selection page}
                \label{fig:UI1}
        \end{subfigure}\hspace{5mm}
        \begin{subfigure}[t]{0.22\textwidth}
                \includegraphics[width=\linewidth]{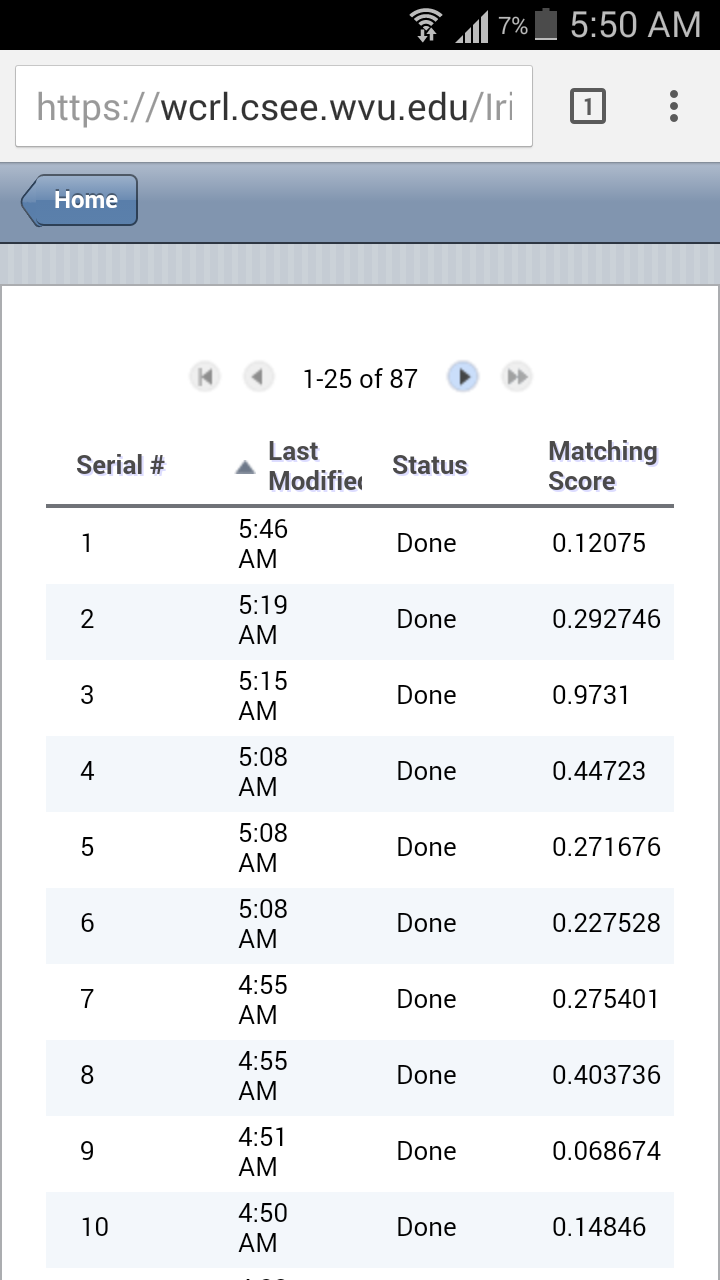}
                \caption{Job history page }
                \label{fig:UI2}
        \end{subfigure}%

\caption{Screen shots of the mobile web app.}\label{fig:UIs} 
\vspace{-0.30cm} 
\end{figure}

3. If the user wishes to submit a new job, they can upload the images and either explicitly select a matching algorithm or allow the interface to select an appropriate algorithm based on image characteristics, as shown in Fig. \ref{fig:UI1}.

4. Upon submitting the job request, the user will be redirected to the Job History Page. Shown in Fig. \ref{fig:UI2} is the Job History page view. This page provides details about all previous jobs submitted by the user. The user can view the complete details of a particular job -- including the input images and the matching score -- by clicking on the associated job. 

\subsection{Developer Interface}

\begin{figure}[t]
\centering
\includegraphics[width=8.2cm]{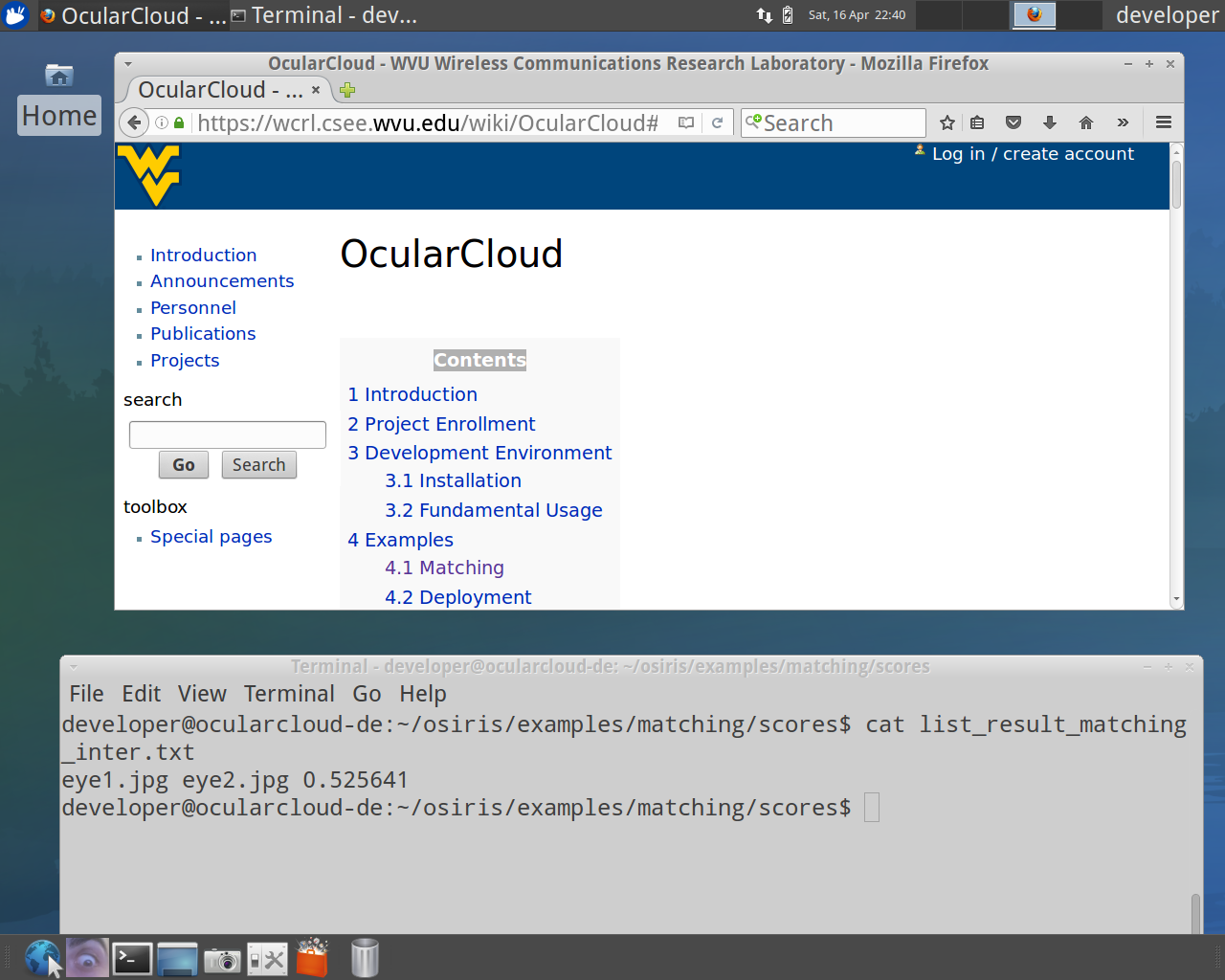}
\caption{Virtual machine implementing the developer interface.  The operating
system is Ubuntu 12.04.  The web browser displays the developer documentation
wiki.  A terminal window shows the match score for two images as computed by OSIRIS.}\label{fig:DI1} 
\vspace{-0.30cm} 
\end{figure}

This subsection describes our implementation of the developer interface based on the model presented in
subsection \ref{subsec:dev_int_model}.
The developer interface is implemented as a virtual machine using \emph{Ubuntu 12.04} (Precise) as the operating
system.
This is the same operating system installed on the cloud infrastructure nodes, which simplifies
the deployment of matching algorithms.
The software tools and libraries used for compiling algorithms in the developer interface
exactly match those on the infrastructure, enabling the developer to deploy
binaries directly.
Virtualbox was chosen as the hypervisor as it is freely available for all major
computing platforms (Windows, OSX, and Linux).
A example of the developer interface is shown in Fig. \ref{fig:DI1}.

The virtual machine is distributed through a publicly accessible
website\footnote[3]{https://wcrl.csee.wvu.edu/wiki/OcularCloud} as a compressed archive which expands to a single Virtualbox Disk Image (VDI)
file.
The developer specifies the VDI file as the disk image for a virtual machine
in Virtualbox, and the developer interface is immediately available.
Downloading and executing a virtual machine image is much simpler than
a conventional provisioning process where the developer personally installs Ubuntu
and the required software.

The virtual machine contains the following software for matching algorithm development:
Open CV \emph{2.4.11} \cite{bradski_opencv_2000}, OSIRIS \emph{4.1}\footnote[4]{http://svnext.it-sudparis.eu/svnview2-eph/ref\_syst//Iris\_Osiris\_v4.1} and standard utilities for
software development in the Linux environment such as GNU Emacs and the GNU compiler collection.
Documentation for using the interface is provided as a wiki, which is linked via the interface desktop.
The developer deploys their algorithm by uploading the required scripts, executables and data files
to their home directory in the cloud and submitting a request for integration to the infrastructure administrators.
Considering OSIRIS, a shell script is provided which automates the algorithm deployment process.

\subsection{Computing Infrastructure}

This subsection describes the implementation of the infrastructure model
described in subsection \ref{subsec:cloudinf_mdl}.
The infrastructure contains twenty-two physical servers having AMD64 architecture,
with a varying number of processing cores and main memory per server.
There are a total of $408$ processing cores and $408$ gigabytes of main memory.
The servers are interconnected via gigabit Ethernet.
A single server acts as a router between the internet and remaining servers,
and hosts the web and shell servers.  This server is denoted as the \emph{head node}.
The remaining servers execute matching requests, and are denoted as
\emph{worker nodes}.
The operating system on all nodes is \emph{Ubuntu Linux} 12.04 (Precise).
All data is stored on the head node and shared to the worker nodes
using the standardized distributed file system protocol \emph{Network File System} (NFS).
The software components implementing the job manager, task controller,
and generic worker are designed to work independently of
one another, communicating through files on the file system.
This architecture allows the components to be re-used with little or no modification
to the code, consistent with the UNIX philosophy and the notion 
of a \emph{microservice} \cite{thones_micro_2015}.


The job manager is implemented as a MATLAB program that is run persistently on the head node 
within a \emph{GNU screen} session.
When a user submits a matching request using the web interface, the interface 
creates a data file (in MATLAB's $.mat$ form) containing (a) the paths to the input ocular
images and (b) the user's algorithm selection option; this file is stored in the user's home directory.
The job manager creates a task file -- also in $.mat$ form -- containing paths to the ocular images and the algorithm to execute.

Like the job manager, the task controller is implemented as a MATLAB program
on the head node that is run in a GNU screen session.
The task controller schedules a user's task for execution when worker node
resources become available.
Exactly one matching request may be executed for every processing core available
on the worker nodes.
Once this limit is reached, further matching requests must wait until
a core becomes available.

A matching request is executed by a generic worker process running on a worker node.
The generic worker process is a MATLAB program which executes the algorithm specified
in the task file.
The task file specifies an \emph{entry function} for the matching algorithm,
which is a MATLAB function implemented by the algorithm developer to
initiate algorithm execution.
Since the algorithm developer has full control of the entry point function,
they may execute a program implemented in any language which can be executed
on the Linux command line by using the MATLAB feature to execute shell commands.

Once the algorithm execution is complete, control is returned to the generic
worker, which stores matching results in the task file.
The task file is consumed by the job manager, which stores the matching
result in the job file.
The job file is passed through the queues to the web interface, which
displays the matching result to the user.

\section{Performance Evaluation}\label{sec:perf}

In this section we describe two sets of experiments that illustrate the potential benefits of the framework. The first experiment shows that different algorithms provide different matching performance, thereby motivating the need for a system that supports a plurality of algorithms. The second experiment evaluates the performance of the system when the algorithm is automatically selected.

\subsection{Matching Performance}
The dataset used for the first experiment is the ND-IRIS-0405 iris dataset \cite{bowyer_ndiris_2010}. This dataset contains 64,980 images corresponding to 356 unique subjects and 712 unique irises. For our evaluation, we use iris images of 12 subjects and 12 images of the same iris per subject. In total, we used 144 images. The matching algorithm used was OSIRIS. The OSIRIS (Open Source for IRIS) is a well known open source iris recognition system developed in the framework of the BioSecure project. It is composed of four processing modules -- segmentation, normalization, encoding, and matching. Gabor filters are applied to the normalized iris image and the resulting phasor responses are quantized into a binary feature set. The Hamming distance measure is used to compare the binary feature sets of two iris images in order to obtain the final matching score. 

Different Gabor Filter parameters were selected for the OSIRIS algorithm, resulting in different sets of Gabor filters. This was accomplished by changing the {\em sizes} of the Gabor filters, but retaining the same {\em number} of Gabor filters. Gabor filter coefficient sizes are defined in terms of the coefficient matrix ($m \times n$). We used three different Gabor filter parameter sets \textbf{A, B, C} for this experiment. The \textbf{A, B, C} parameter sets have $2$ Gabor filters each, with coefficient matrix sizes of $9 \times 15$, $9 \times 27$ and $9 \times 51$, respectively. Each parameter set is viewed as a different OSIRIS algorithm.  

\begin{figure}[t]
\centering
\includegraphics[width=8.85 cm]{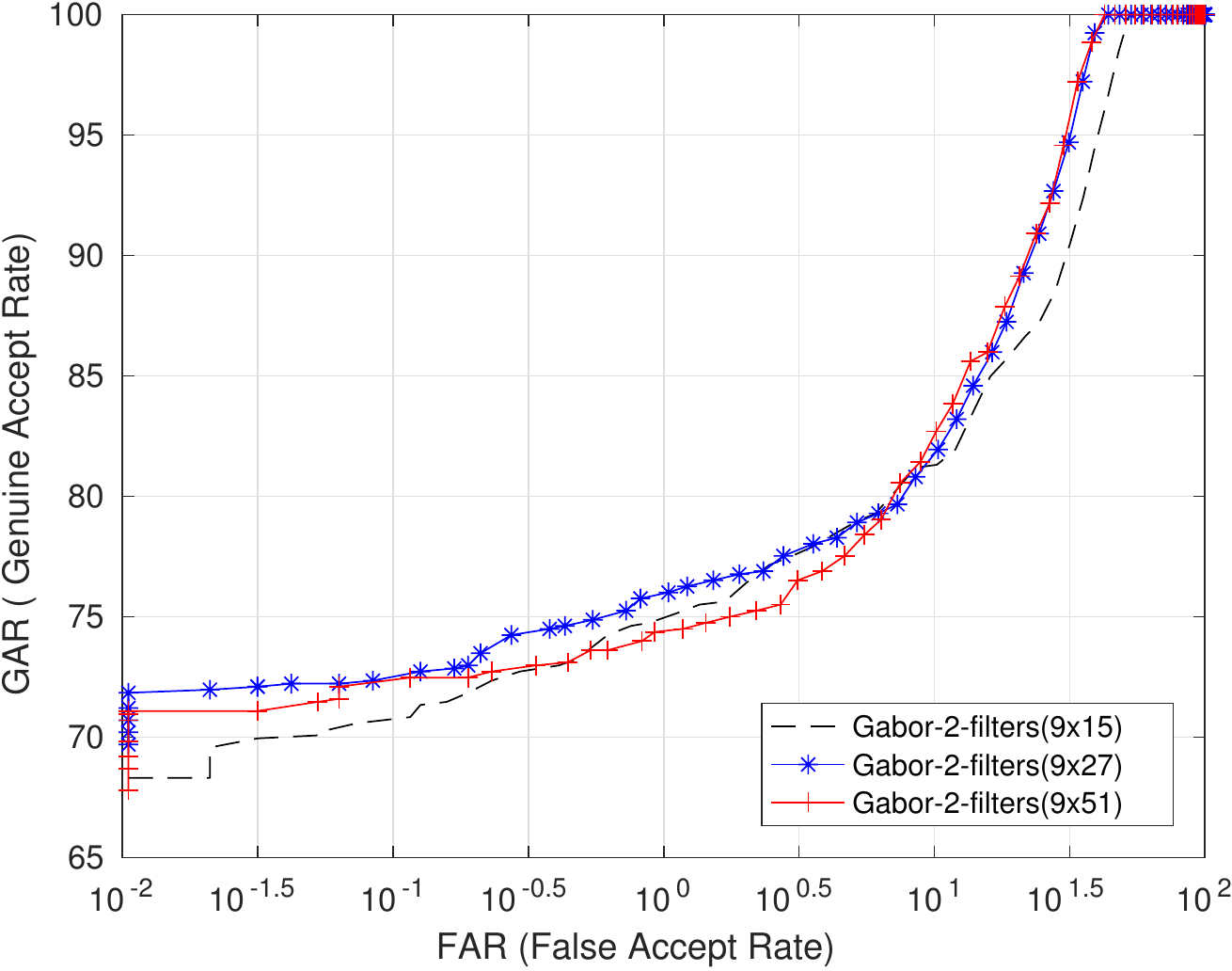}
\caption{ROC curve for same number of Gabor filters but different sizes.}\label{fig:ROC1} 
\vspace{-0.30cm}
\end{figure}

The following experiment was performed to evaluate and compare the performance of these three algorithms. False Accept Rate (FAR), False Reject Rate (FRR) and Genuine Accept Rate (GAR) are computed for the test data set of $144$ images. Based on the number of subjects ($N=12$) and the number of images ($t=12$) per subject we obtain $Nt(t-1)/2 = 792$ genuine scores and $(N(N-1)t^2)/2=9504$ imposter scores. The ROC (GAR vs FAR) curve at various threshold points for the first experiment is shown in Fig. \ref{fig:ROC1}. The Gabor filters of size $9 \times 27$ perform marginally better than the other two algorithms. But it can be observed from the curves that there is no clear winner. However, these curves suggest that different algorithms may be needed depending upon operational requirements of FAR and/or GAR.  

\subsection{Automatic Selection}

In order to evaluate the BaaS framework, we conducted experiments by using the web app on a smartphone as well as on a personal computer (PC). The smartphone experiments were conducted using a live image of the ocular region (taken using the front or the back camera of the phone) as one input, and an ocular image from the phone gallery as the second input. The PC experiments were conducted using images from the ND-IRIS-0405 iris data-set \cite{bowyer_ndiris_2010}. 220 tests (each test entailed matching two images)  were conducted using smartphone and 100 tests were conducted using the PC. The tests consisted of both genuine and impostor image pairings.  

Once the images were uploaded using the smartphone or a PC, the algorithm to be executed on the input images was selected automatically. Based on the input image characteristics, a particular algorithm was automatically invoked by the system at the time of authentication. In this experiment, two types of algorithms were considered. The first was an OSIRIS-based iris recognition algorithm and the other was a custom periocular matching algorithm. The selection method first computes the radius of the iris region in the input ocular image and uses this to select one of the two algorithms.  In particular, if the radius of the iris is below a threshold the algorithm ``Periocular'' is selected; otherwise the ``OSIRIS'' algorithm is selected for execution.

\begin{table}[t]
\centering
\captionsetup{width=.9\linewidth}
\caption{Table showing the number of executions for each algorithm for a total of 320 tests}
\label{table:1}
\begin{tabular}{||m{0.5cm}|m{1.00cm}|m{1.00cm}|m{1.00cm}|m{1.00cm}|m{1.00cm}||} 
 \hline
 Serial No. & Developer Name & Algorithm Name & Modality & Number of Executions & Number of Executions in \% \\ [0.5ex] 
 \hline\hline
 1 & TMPS & OSIRIS & Iris & 246 & 76.9\% \\ [0.5ex]
 \hline
 2 & ROSS & Periocular & Ocular & 74 & 23.1\% \\ [0.5ex] 
 \hline
\end{tabular}

\vspace{-0.30cm}
\end{table}

Table \ref{table:1} gives the number of times each algorithm was executed during the 320 tests conducted in this work.  Besides hosting two completely different algorithms, it is possible for the cloud to host several instances of the same algorithm, where the different instances use different parameters. The experiment conveys the main theme of the proposed framework, i.e., depending on the input images, a different algorithm is selected each time and the developer for that selected algorithm is rendered a micropayment. This experiment shows that the proposed framework is feasible and creates an innovative and competitive ecosystem that benefits both software developers and end-users.

\section{Summary}\label{sec:concl}

In this paper, we proposed a framework for performing biometric matching in a cloud environment using the sensors available in ordinary smartphones.  The proposed biometrics-as-a-service paradigm enables users to perform biometric matching in a web interface.  Moreover, the platform-as-a-service model enables the developers of recognition technology to upload their algorithms to the cloud. By selecting algorithms for execution and rendering micropayments to the corresponding developer, continuous innovation is encouraged. An implementation has been developed demonstrating that the architecture is feasible in the form of a case study based on ocular recognition. Future work will be directed towards optimal matching algorithm selection and using concepts from game theory to determine optimal pricing points.

\section*{Acknowledgment}

This research was funded by the Center for Identification Technology Research (CITeR), a National Science Foundation (NSF) Industry/University Cooperative Research Center (I/UCRC).

\balance
\vspace{-0.35cm}
\bibliographystyle{IEEEtran}
\bibliography{milcom2016}

\end{document}